\documentclass{Interspeech}

\usepackage{multirow}
\usepackage{amsmath,graphicx}
\usepackage[subrefformat=parens]{subcaption}
\usepackage{cite}

\interspeechcameraready

\title{Prosody Labeling with Phoneme-BERT and Speech Foundation Models}
\author[affiliation=1]{Tomoki}{Koriyama}

\affiliation{}{CyberAgent}{Japan}
\email{koriyama\_tomoki@cyberagent.co.jp}
\keywords{prosody, pitch accent labels, BERT, self-supervised-learning models, pretrained models}

\usepackage{comment}

\begin{document}

\maketitle
\begin{abstract}

This paper proposes a model for automatic prosodic label annotation, where the predicted labels can be used for training a prosody-controllable text-to-speech model.
The proposed model utilizes not only rich acoustic features extracted by a self-supervised-learning (SSL)-based model or a Whisper encoder, but also linguistic features obtained from phoneme-input pretrained linguistic foundation models such as PnG BERT and PL-BERT.
The concatenation of acoustic and linguistic features is used to predict phoneme-level prosodic labels.
In the experimental evaluation on Japanese prosodic labels, including pitch accents and phrase break indices,
it was observed that the combination of both speech and linguistic foundation models enhanced the prediction accuracy compared to using either a speech or linguistic input alone. 
Specifically, we achieved 89.8\% prediction accuracy in accent labels, 93.2\% in high-low pitch accents, and 94.3\% in break indices.

\end{abstract}

\section{Introduction}

Recent text-to-speech (TTS) technology has enabled us to generate human-like utterances. To achieve more human-like speech synthesis, it is necessary to accurately produce not only phonetic information but also prosody. For example, syllable-level prosody such as tones and stress/pitch accents is used to distinguish the different meanings of words. Boundary pitch movements uttered with rising pitch are often the expression of interrogative sentences. Phrase boundaries are also important to represent the semantic relationship among words.

One approach to achieving TTS with rich prosodic expression is to introduce an implicit intermediate representation of prosody \cite{ogura2025mora, koriyama2019semisupervised, yufune2021accent, yamauchi2024cross, zhong2024multimodal, liu2024language}. Specifically, they proposed pitch accent representation by 
mora-level fundamental frequency (F0) \cite{ogura2025mora} or accent latent variables \cite{koriyama2019semisupervised, yufune2021accent, yamauchi2024cross}. 
Zhang et al. \cite{zhong2024multimodal} proposed phrase boundary representations obtained from the similarity of speech-silence and word-punctuation. By using intermediate representations and predicting them from input text, we can generate prosody without using prosodic labels for training.

However, this approach makes it difficult to control prosody if the generated prosody is not what the user intends.
Hence, another approach is to employ explicit prosodic labels in the training of TTS, where the labels are estimated by automatic prosodic annotation \cite{dai2022automatic, shirahata2024audio, kurihara2024low, vetter2019crosslingual, park2022unified, zhai2023wav2tobi}.
Dai et al. \cite{dai2022automatic} proposed a phrase break prediction model using an acoustic encoder based on Conformer \cite{gulati20conformer} and a linguistic encoder based on Bidirectional Encoder Representation From Transformer (BERT) \cite{devlin2019bert}.
Shirahata et al. \cite{shirahata2024audio} achieved a simultaneous prediction model of accent symbols based on Whisper \cite{radford2023robust}-based acoustic feature extraction and automatic speech recognition (ASR) architecture.
A similar approach was performed by Kurihara and Sano \cite{kurihara2024low}, in which wav2vec2.0 \cite{baevski2020wav2vec} was used for acoustic feature extraction.
Although automatic prosodic annotation requires a certain amount of labeled training data, we can control generated prosody by providing explicit labels.
For example, we can manually correct errors in the accents of proper nouns, such as peoples' names and place names, and fix phrase breaks that might cause misunderstandings.
In this study, we focus on this automatic prosody annotation.

In recent years, pre-trained foundation models trained on large datasets have attracted attention in the field of prosody modeling. Since foundation models are already trained on a large amount of data, they are useful for effective feature extraction even if text-speech-paired data is limited.
For example, the linguistic foundation model based on BERT and its alternatives enhanced the naturalness of prosody \cite{hayashi2019pretrained, xiao2020improving, kenter2020improving, xu2021improving, zhong2024multimodal, yasuda2019investigation, ogura2025mora, liu2024pewav2vec} due to the rich representation of linguistic features.
Moreover, phoneme-input BERT models such as PnG BERT \cite{jia2021png} and PL-BERT \cite{li2023phoneme} enabled us to express phoneme-level prosody. 
As for speech foundation models, the Whisper encoder \cite{radford2023robust} was effective for extracting acoustic features that represent prosodic information \cite{shirahata2024audio}. Self-supervised learning (SSL) models such as wav2vec2.0  has also been shown to be effective in prosody modeling \cite{kunesova2022detection, liu2024language, kurihara2024low}. Furthermore, Yamauchi et al. demonstrated that it is possible to estimate the latent variables obtained from the Whisper encoder using PL-BERT \cite{yamauchi2024cross}.

This gives rise to a question: what if we combine linguistic and speech foundation models in automatic prosody annotation? Hence, we propose a prediction model of phoneme-level prosodic labels, in which the phoneme-level inputs are given by the concatenation of linguistic and speech foundation model outputs.
Due to the proposed model, we expect to capture not only word-dependent prosodic boundaries using linguistic models but also prosodic pitch movements from acoustic models.
In the experimental evaluations, the proposed model was trained using the corpus of spontaneous Japanese (CSJ) \cite{Maekawa_2003}, including manually-annotated rich prosodic labels. PnG BERT \cite{jia2021png} and PL-BERT \cite{li2023phoneme} were used as linguistic foundation models. SSL models such as HuBERT \cite{hsu2021hubert}, wav2vec2.0 \cite{baevski2020wav2vec}, and WavLM \cite{chen2022wavlm} and Whisper \cite{radford2023robust} were used as speech foundation models. Then, we investigated the effectiveness of combining linguistic and speech models.
We used multiple types of prosodic labels composed of accent symbols, high-low symbols, break indices, and pause presence as the target of prediction.
Experimental results show that the proposed combination is more effective than using either linguistic or speech foundation models alone and outperforms traditional acoustic features such as melspectrogram and F0.

\section{Related Work}

Shirahata et al. \cite{shirahata2024audio} have also conducted automatic prosody annotation for constructing a speech synthesis database, similar to the objective of our study.
Their annotation model consists of a Whisper-based encoder-decoder architecture, with a pretrained Whisper encoder for feature extraction and a decoder that predicts a mixed sequence of phonemes and accent symbols.
It is true that for the creation of a speech synthesis database, it is valuable to recognize both phonemes and accent symbols simultaneously. However, phoneme recognition and pitch-accent recognition are problems of different levels of difficulty. Phonemes, or pronunciations, can be annotated by many people, enabling training with a large amount of data. On the other hand, prosody annotation can only be performed by a limited number of professionals, requiring training with a limited amount of data. Therefore, we treat phoneme recognition and accent recognition as separate problems. In this study, we assume that phonemes can be obtained from transcription information or other labels and propose a prosody annotation system that uses phonemes as input.

\begin{figure}[t]
  \centering
  \includegraphics[width=0.9\linewidth]{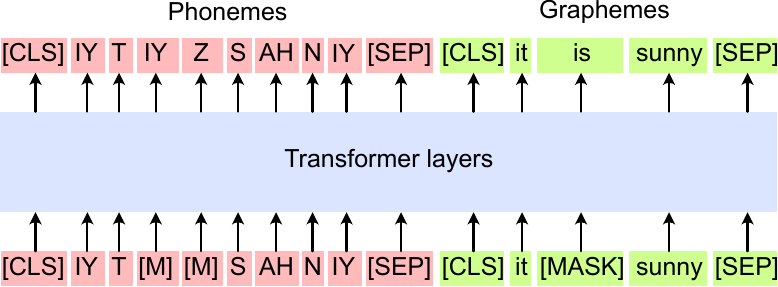}
  \caption{Training scheme of PnG BERT. Masked phonemes [M] and masked token [MASK] are inferred.}
  \label{fig:png_bert}
\end{figure}

\begin{figure}[t]
  \centering
  \includegraphics[width=0.9\linewidth]{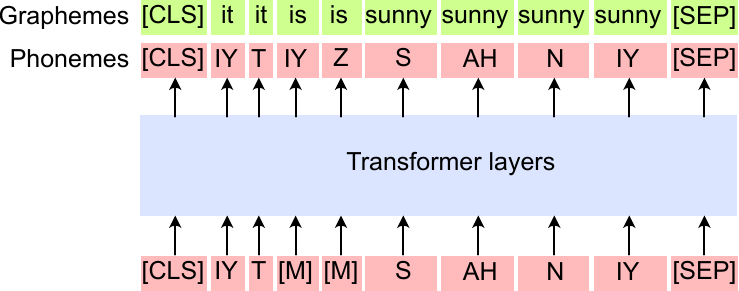}
  \caption{Training scheme of PL BERT. Masked phonemes [M] and the target word tokens are inferred.}
  \label{fig:pl_bert}
\end{figure}

\section{Linguistic and Speech Foundation Models}

\subsection{Linguistic foundation models with phoneme inputs}
SSL-based linguistic models such as BERT have been incorporated into TTS models and have made it possible to improve the synthetic speech quality due to the rich linguistic information extraction \cite{hayashi2019pretrained, xiao2020improving, kenter2020improving, xu2021improving}.
Following the success of BERT, extended BERT models that can input phonemes have been proposed for phoneme-input TTS frameworks that can generate utterances with correct pronunciations.
PnG BERT \cite{jia2021png}, shown in Fig.~\ref{fig:png_bert}, is a BERT model that inputs a concatenated sequence of phonemes and graphemes. In the training of PnG BERT, words and their corresponding phonemes are simultaneously masked, and the transformer layer parameters are optimized for predicting the masked tokens.
It has been shown that the use of PnG BERT enhances the naturalness of synthetic speech by utilizing BERT's feature extraction capabilities.

Another phoneme-input BERT is phoneme-level BERT (PL-BERT) \cite{li2023phoneme}, shown in Fig.~\ref{fig:pl_bert}. Since PnG BERT uses graphemes as inputs, it cannot handle unseen tokens. To overcome this problem, PL-BERT uses phonemes alone as inputs, while graphemes are used as the target in training.
It has been shown that PL-BERT also outperformed TTS models without BERT encoders.

\begin{figure}[t]
  \centering
  \includegraphics[width=\linewidth]{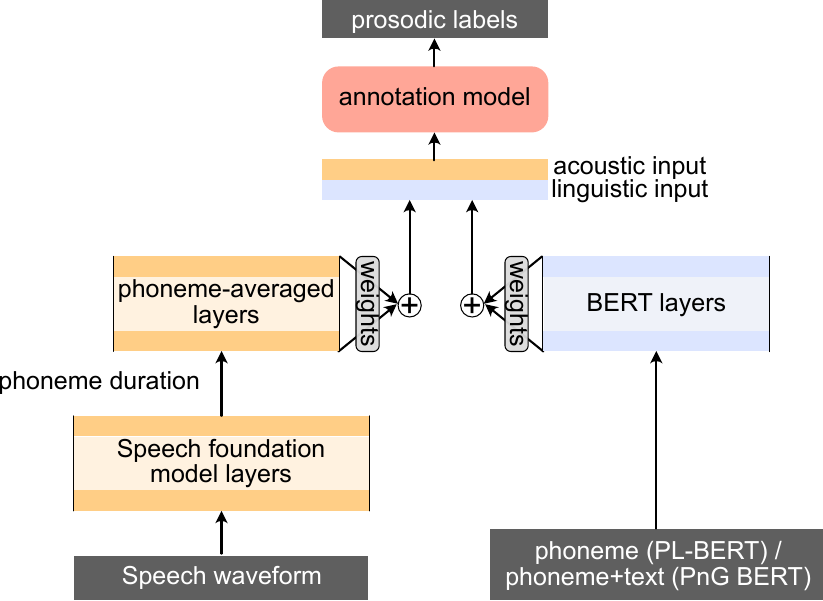}
  \caption{Proposed model architecture.}
  \label{fig:model}
\end{figure}

\subsection{Speech Foundation Models}

In speech models, self-supervised learning inspired by BERT has also been shown to be effective for speech feature extraction.
Popular models include wav2vec2.0 \cite{baevski2020wav2vec}, which is trained by contrastive learning, and HuBERT \cite{hsu2021hubert}, which uses clustering centroids as the targets.
WavLM \cite{chen2022wavlm} is an extension of HuBERT that employs noisy audio inputs to handle diverse types of speech.
It has been shown that the SSL models can be used for various applications, including ASR and emotion classification \cite{mohamed2022self, li2023styletts2}.

Whisper \cite{radford2023robust} is an ASR model trained on a large dataset.
Although Whisper itself is an encoder-decoder-based model,
its modules can be used as a feature extractor for diverse applications \cite{shirahata2024audio, ma2024emobox, mogridge2024nonintrusive}.

\begin{figure}[t]
  \centering
  \includegraphics[width=0.95\linewidth]{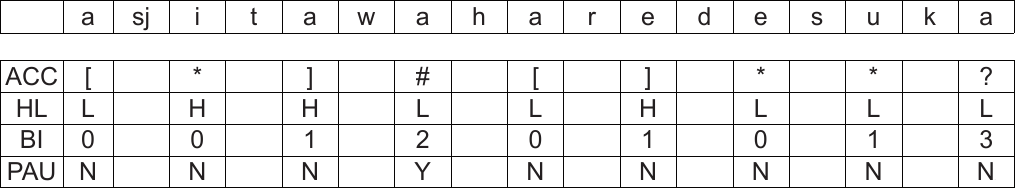}
  \caption{Multi-level prosody labels used for this study. The sentence is ``ashita wa hare desuka'' in Japanese that means ``Will it  be sunny tomorrow?''}
  \label{fig:labels}
\end{figure}

\begin{table*}[t]
  \caption{Prediction accuracies on the combination of acoustic and linguistic models.}
  \footnotesize
  \label{table:combination}
  \centering
  \begin{tabular}{ ll| rrrr | rrrr }
    \toprule
     & & \multicolumn{4}{c|}{Accuracy} & \multicolumn{4}{c}{Macro F1} \\
    Aco. model & Ling. model & ACC & HL & BI & PAU & ACC & HL & BI & PAU \\
    \midrule
HuBERT-base  &  PnG BERT  &  \textbf{0.898}  &  \textbf{0.932}  &  \textbf{0.943}  &  0.987  &  \textbf{0.852}  &  \textbf{0.932}  &  0.876  &  0.820 \\
  &  PL-BERT  &  0.894  &  0.931  &  0.937  &  \textbf{0.988}  &  0.842  &  0.931  &  0.866  &  \textbf{0.828} \\
  &  One-hot  &  0.892  &  0.930  &  0.929  &  0.987  &  0.847  &  0.930  &  0.855  &  0.815 \\
  &  None  &  0.890  &  0.928  &  0.928  &  0.987  &  0.842  &  0.928  &  0.851  &  0.819 \\
    \midrule
Whisper-small  &  PnG BERT  &  0.892  &  0.928  &  0.942  &  \textbf{0.988}  &  0.845  &  0.928  &  \textbf{0.878}  &  0.819 \\
  &  PL-BERT  &  0.890  &  0.929  &  0.936  &  \textbf{0.988}  &  0.844  &  0.929  &  0.863  &  0.817 \\
  &  One-hot  &  0.885  &  0.922  &  0.921  &  0.987  &  0.834  &  0.922  &  0.847  &  0.816 \\
  &  None  &  0.883  &  0.921  &  0.919  &  0.987  &  0.838  &  0.921  &  0.839  &  0.801 \\
    \midrule
Melspec  &  PnG BERT  &  0.854  &  0.898  &  0.925  &  0.983  &  0.713  &  0.898  &  0.835  &  0.701 \\
  &  PL-BERT  &  0.847  &  0.892  &  0.914  &  0.983  &  0.704  &  0.892  &  0.813  &  0.694 \\
  &  One-hot  &  0.821  &  0.871  &  0.865  &  0.983  &  0.680  &  0.871  &  0.748  &  0.684 \\
  &  None  &  0.752  &  0.830  &  0.743  &  0.981  &  0.598  &  0.830  &  0.593  &  0.605 \\
    \midrule
F0  &  PnG BERT  &  0.854  &  0.896  &  0.929  &  0.984  &  0.700  &  0.896  &  0.842  &  0.677 \\
  &  PL-BERT  &  0.849  &  0.891  &  0.918  &  0.984  &  0.698  &  0.891  &  0.812  &  0.667 \\
  &  One-hot  &  0.837  &  0.879  &  0.900  &  0.984  &  0.695  &  0.879  &  0.770  &  0.660 \\
  &  None  &  0.629  &  0.736  &  0.573  &  0.981  &  0.402  &  0.735  &  0.381  &  0.506 \\
    \midrule
None  &  PnG BERT  &  0.825  &  0.870  &  0.914  &  0.983  &  0.646  &  0.870  &  0.821  &  0.634 \\
  &  PL-BERT  &  0.817  &  0.863  &  0.902  &  0.983  &  0.643  &  0.863  &  0.782  &  0.634 \\
  &  One-hot  &  0.799  &  0.841  &  0.886  &  0.983  &  0.612  &  0.840  &  0.752  &  0.604 \\
  &  None  &  -  &  -  &  -  &  -  &  -  &  -  &  -  &  - \\
     \bottomrule
  \end{tabular}
\end{table*}

\begin{figure*}[t]
  \centering
  \includegraphics[width=0.8\linewidth]{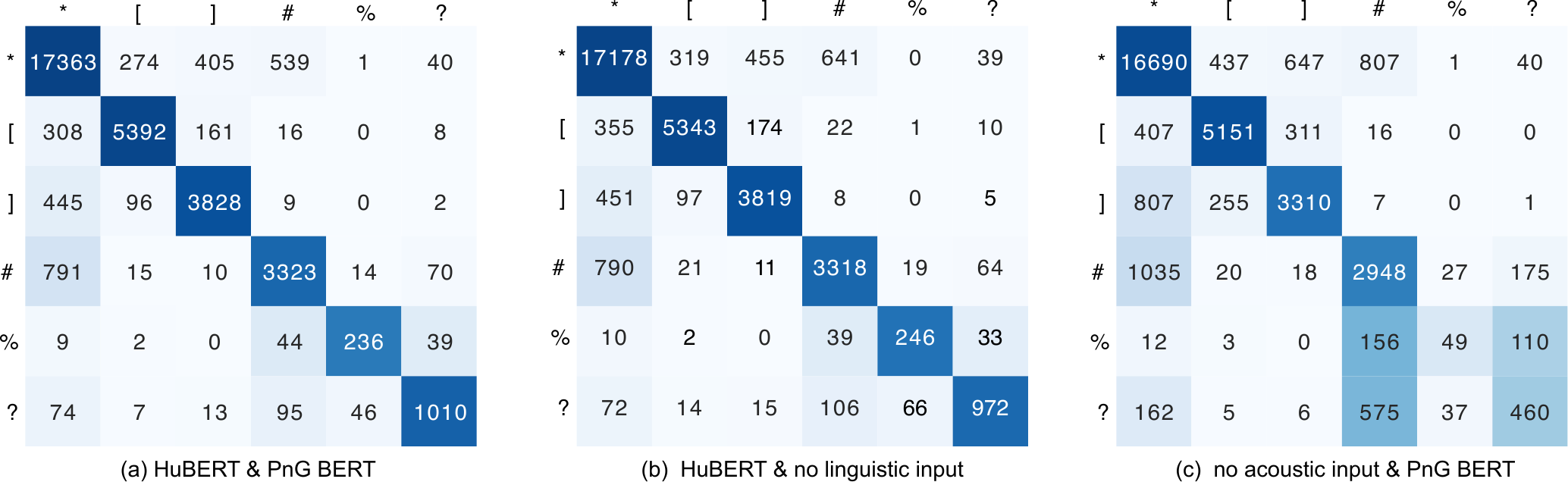}
  \caption{Confusion matrices on accent symbol (ACC) prediction.}
  \label{fig:confusion_acc}

  \vspace{10pt}
  \centering
  \includegraphics[width=0.5\linewidth]{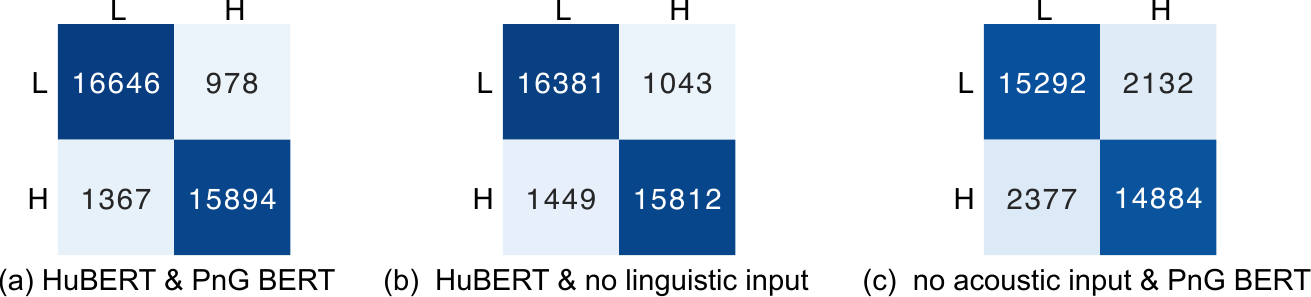}
  \caption{Confusion matrices on high-low (HL) prediction.}
  \label{fig:confusion_hl}
  \vspace{10pt}

  \centering
  \includegraphics[width=0.8\linewidth]{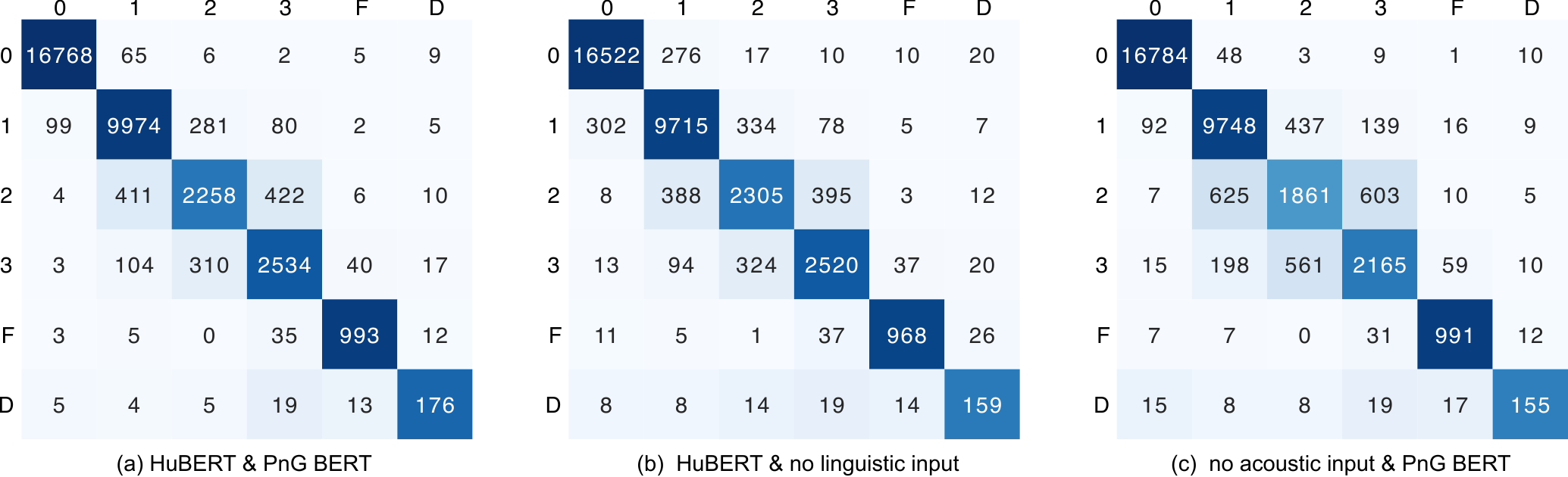}
  \caption{Confusion matrices on break index (BI) prediction.}
  \label{fig:confusion_bi}

  \vspace{10pt}
  \centering
  \includegraphics[width=0.5\linewidth]{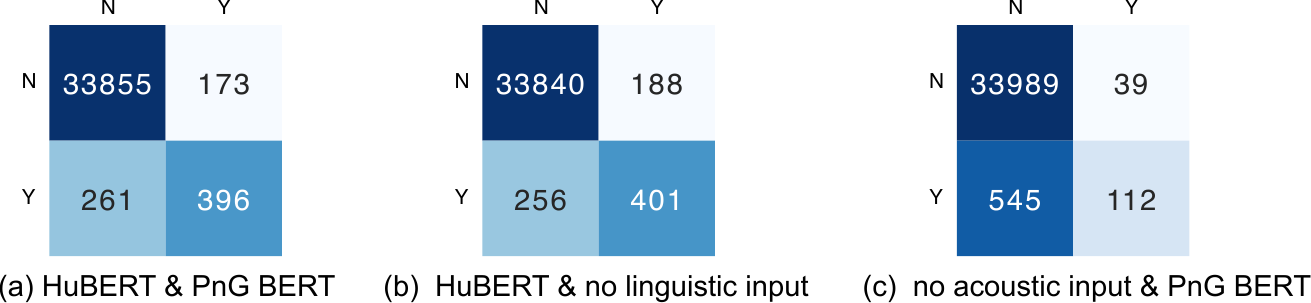}
  \caption{Confusion matrices on pause presence (PAU) prediction.}
  \label{fig:confusion_pau}
\end{figure*}

\section{Proposed prosody annotation model}

We propose an automatic prosody annotation model
using speech and linguistic foundation models.
Figure~\ref{fig:model} shows the outline of the proposed model.
A speech waveform is input into a speech foundation model such as an SSL encoder or Whisper encoder, and the acoustic features at a frame level are obtained from the encoder's hidden layers.
By injecting the phoneme durations given by phoneme alignment modules \cite{mcauliffe2017montreal, badlani2022one, koriyama2025vae} and averaging the frame-level features,  
we compute phoneme-level acoustic features.
Moreover, we extract phoneme-level linguistic features using linguistic foundation models, including PnG BERT and PL-BERT.
After calculating the weighted sums of hidden layers of extracted acoustic and linguistic features, respectively, we input the concatenated acoustic and linguistic features into an annotation model to predict phoneme-level prosodic labels.
In this way, we can utilize both waveform- and text-based information for prosody label prediction.

For the training of the proposed model,
we freeze the parameters of speech and linguistic foundation models
and optimize the weights for layer selection and the parameters of the annotation model.
We can use multi-level prosodic labels such as tones/stresses for syllables and break indices for phrases.
In that case, we use multi-task learning for the training.

Traditional acoustic features such as MFCC and melspectrogram are hand-crafted. Although they are effective in speech processing tasks, recent works have proven that large-data-driven features, such as SSL acoustic features, are more effective than traditional ones \cite{mohamed2022self}. Therefore, we can also expect effectiveness in the prosody annotation task.
Additionally, while F0 is a crucial acoustic feature for capturing pitch contour, its extraction can be challenging in noisy environments, often resulting in inaccurate pitch contours.

As for linguistic foundation models, 
we expect that they can capture the specific functions of phoneme sequences. For instance, in English, the syllables that are likely to have stress can be obtained from phoneme sequences, and in Thai, syllable-level stress that determines tone expression intensities may depend on linguistic inputs \cite{peyasantiwong1986stress}.


\section{Experiments}

\subsection{Pretraining of phoneme-input BERT models}

We pre-trained language models of PnG BERT and PL-BERT. 
We used Japanese Wikipedia corpus \cite{wikipedia_corpus} containing 4.9 GB and the Japanese part of the CC-100 dataset \cite{cc100} consisting of 70 GB sentences.
The total number of sentences were approximately 630 millions.

In the original PnG BERT and PL-BERT, word-level graphemes were used for training. However, due to the uncertainty of word boundaries and the vast vocabulary size in Japanese, in this study, we used subwords as the grapheme inputs. Specifically, we first performed word segmentation using UniDic-lite \cite{unidic} implemented by fugashi\footnote{\url{https://github.com/polm/fugashi}}, followed by subword segmentation using a tokenizer implemented by \textit{cl-tohoku/bert-base-japanese-v3}\footnote{\url{https://huggingface.co/tohoku-nlp/bert-base-japanese-v3}}.
Phonemes were obtained by the pronunciation dictionary of Unidic.
The resulting phoneme-subword alignments (which phoneme and subword corresponds to which word, respectively) were used for training PnG BERT and PL-BERT.
Different from the original PnG BERT, we did not input word position indices to avoid the phoneme-subword alignment in inference.

The vocabulary size was 32,768, and the number of phonemes was 62, based on the phoneme set of CSJ \cite{Maekawa_2003}.
The hidden size was 768 and the number of hidden layers was 12.
In the training of both of PnG BERT and PL-BERT, the minibatch size was 64 and the parameters were updated by 1 M steps. An Adam optimizer \cite{kingma2014adam} with a learning rate $1.0 \times 10^{-5}$ and a constant scheduler with 500 warm-up steps were used.

\begin{table*}[t]
  \caption{Prediction accuracies on SSL-based acoustic models. PnG BERT is used as the linguistic foundation model. Model names are those at \url{huggingface.co}.}
  \footnotesize
  \label{table:ssl_models}
  \centering
  \begin{tabular}{ lcc | rrrr | rrrr }
    \toprule
     & & & \multicolumn{4}{c|}{Accuracy} & \multicolumn{4}{c}{Macro F1} \\
    Aco. model & Language & \# layers & ACC & HL & BI & PAU & ACC & HL & BI & PAU \\
    \midrule
rinna/japanese-hubert-base & Japanese & 12 & 0.898 & 0.932 & 0.943 & 0.987 & \textbf{0.852} & 0.932 & 0.876 & 0.820 \\
rinna/japanese-hubert-large & Japanese & 24 & 0.894 & 0.929 & 0.943 & \textbf{0.988} & 0.845 & 0.929 & 0.878 & 0.816 \\
rinna/japanese-wav2vec-base & Japanese & 12 & \textbf{0.897} & \textbf{0.934} & \textbf{0.944} & \textbf{0.988} & 0.851 & \textbf{0.934} & \textbf{0.880} & \textbf{0.825} \\
    \midrule
facebook/hubert-base-ls960 & English & 12 & 0.888 & 0.922 & 0.938 & 0.987 & 0.833 & 0.922 & 0.865 & 0.812 \\
facebook/hubert-large-ll60k & English & 24 & 0.886 & 0.921 & 0.941 & \textbf{0.988} & 0.830 & 0.921 & 0.871 & 0.799 \\
facebook/wav2vec2-base & English & 12 & 0.889 & 0.927 & 0.938 & 0.987 & 0.834 & 0.927 & 0.867 & 0.812 \\
facebook/wav2vec2-large & English & 24 & 0.891 & 0.925 & 0.941 & 0.987 & 0.844 & 0.925 & 0.868 & 0.794 \\
    \midrule
facebook/wav2vec2-large-xlsr-53 & Multiple & 24 & 0.888 & 0.923 & 0.940 & \textbf{0.988} & 0.833 & 0.923 & 0.872 & 0.807 \\
microsoft/wavlm-base-plus & Multiple & 12 & 0.890 & 0.925 & 0.941 & 0.987 & 0.842 & 0.925 & 0.874 & 0.811 \\
microsoft/wavlm-large & Multiple & 24 & 0.892 & 0.927 & 0.943 & \textbf{0.988} & 0.833 & 0.923 & 0.872 & 0.807 \\
\bottomrule
  \end{tabular}
\vspace{10pt}

  \caption{Prediction accuracies on Whisper-based acoustic models. PnG BERT is used as the linguistic foundation model.}
  \footnotesize
  \label{table:whisper}
  \centering
  \begin{tabular}{ lc | rrrr | rrrr }
    \toprule
      & &\multicolumn{4}{c|}{Accuracy} & \multicolumn{4}{c}{Macro F1} \\
    Aco. model  & \# layers & ACC & HL & BI & PAU & ACC & HL & BI & PAU \\
    \midrule
Whisper-tiny &	4 & 0.892 &	0.927 &	0.941 &	0.987 &	0.840 &	0.927 &	0.873 &	0.805 \\
Whisper-base & 6 & 0.890 &	0.924 &	0.943 &	0.987 &	0.839 &	0.924 &	0.878 &	0.811 \\
Whisper-small	& 12 &	0.892 &	0.928 &	0.942 &	\textbf{0.988} &	0.845 &	0.928 &	0.878 &	0.819 \\
Whisper-medium	& 24 &	0.893 &	0.929 & 0.943 &	\textbf{0.988} &	0.842 &	0.929 &	0.877 &	0.807 \\
Whisper-large	& 32 &	0.893 &	0.930 &	0.944 &	\textbf{0.988} &	0.846 &	0.930 &	0.880 &	0.822 \\
Whisper-large-v2 & 32 &	0.893 &	0.927 &	0.944 &	\textbf{0.988} &	\textbf{0.851} &	0.927 &	0.880 &	\textbf{0.826} \\
Whisper-large-v3 & 32 &	\textbf{0.897} &	\textbf{0.931} &	\textbf{0.945} &	0.987 &	\textbf{0.851} &	\textbf{0.931} &	\textbf{0.881} &	0.819 \\
     \bottomrule
  \end{tabular}
\end{table*}

\subsection{Prosody labels}

We used four types of prosody labels based on the intonation annotation of CSJ.
All labels were mora-level ones as shown in Fig.~\ref{fig:labels}.
\begin{description}
    \item[Accent symbols (ACC):] This notation was based on end-to-end TTS symbols introduced by Kurihara et al. \cite{kurihara2021prosodic}. To express high-low pitch accents of Japanese,
    the transition points between high and low moras were denoted by accent symbols. Moreover, the accent phrase boundaries were represented by accent symbols, and boundary pitch movements such as those in an interrogative sentence were also distinguished by the symbols. Specifically, the following labels were used.
    \begin{itemize}
    \item ``*'': \textit{Other} symbol that was neither a high-low transition nor a phrase boundary.
    \item ``['': Transition from low-pitch mora to high-pitch one.
    \item ``]'': Transition from high-pitch mora to low-pitch one.
    \item ``\#'': Accent phrase boundary with fall (normal) pitch movement.
    \item ``\%'': Accent phrase boundary with rise-fall pitch movement, often appearing in speech acts such as confirmation and turn-taking continuation.
    \item ``?'': Accent phrase boundary with rise (including fall-rise) pitch movement, often appearing in questions.
    \end{itemize}

    \item[High-Low (HL):] The pitch accents can be represented by high-low expression. We used mora-level ``high'' and ``low,'' which would be more direct expression of mora-level acoustic characteristics than the transitions of high and low moras of ACC labels.
    \begin{itemize}
    \item ``L'': Low-pitch accent at the corresponding mora.
    \item ``H'': High-pitch accent at the corresponding mora.
    \end{itemize}

    \item[Break Index (BI):] This label represents the phrase boundary intensity, i.e., how the prosody of the input utterance is hierarchically structured. This label is based on the break indices of X-JToBI \cite{Maekawa_2002}.
    \begin{itemize}
    \item ``0'': No boundary.
    \item ``1'': Short-unit word boundary.
    \item ``2'': Accentual phrase boundary. One accent phrase includes no or one accent nucleus.
    \item ``3'': Intonation phrase boundary, which is annotated where a pitch reset occurs.
    \item ``F'': The end of a filled-pause phrase.
    \item ``D'': The end of a disfluency phrase that occurs in word fragments.
    \end{itemize}

    \item[Pause presence (PAU):] In CSJ, short pauses are annotated as the additional information of phrase breaks,
    which are annotated like ``2+p'' with break indices.
    We used PAU labels as the presence of short pauses.
    \begin{itemize}
    \item ``N'': No short pause.
    \item ``Y'': The presence of a short pause.
    \end{itemize}
    
\end{description}
We used all above labels as the target of the proposed model, and cross-entropy losses were calculated individually for each label category and summed up.
Since the labels were mora-level, only mora-core phonemes, such as vowels, a double consonant /Q/, and a syllabic nasal /N/, were used for calculating the losses, and the predicted labels for other phonemes such as consonants were ignored.

\begin{figure*}[t]
  \centering
  \includegraphics[width=0.9\linewidth]{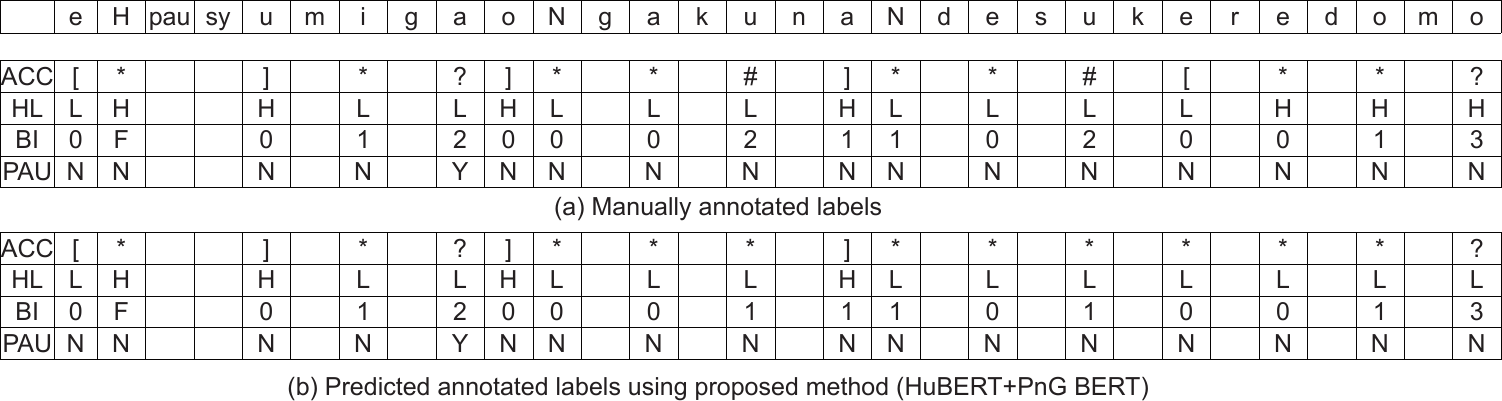}
  \caption{An example of automatically annotated labels. The
sentence is ``ee shumi ga ongaku nandesu keredomo” in Japanese that means
``Well, my hobby is music.''}
  \label{fig:predicted_labels}
\end{figure*}

\subsection{Experimental conditions}

For experimental evaluations, we used the core data of CSJ \cite{Maekawa_2003}, which includes manually annotated prosodic labels.
We divided the speech utterances into training, development, and evaluation subsets
based on the ASR recipe of ESPnet \cite{watanabe2018espnet}.
The training, development, and evaluation subsets consisted of 
24,863 utterances (188 talks,  \textasciitilde 33.2 h),
900 utterances (8 talks, \textasciitilde 1.2 h),
and 896 utterances (8 talks, \textasciitilde 1.3 h),
respectively.
X-JToBI labels were converted into the mora-level prosodic labels.
We used phoneme annotations in CSJ for the phoneme inputs.
The phoneme durations for calculating phoneme-averaged acoustic features were computed by VAE-based alignment results \cite{koriyama2025vae}.

The annotation model was composed of 6-layer convolutional neural networks with a hidden size of 256 and a kernel size of 5.
The training was performed up to 100 k steps using Adam optimization \cite{kingma2014adam} with a learning rate of $1.0 \times 10^{-5}$.
The minibatch size was 4.

For the base SSL model except in Sect.~\ref{sect:aco_feat}, we used the HuBERT-base model trained with the Japanese ReazonSpeech \cite{reazonspeech} database \cite{rinna_hubert}.
For the Whisper encoder except in Sect.~\ref{sect:aco_feat}, Whisper-small was used. Both the HuBERT-base and Whisper-small encoder had a hidden size of 768 and 12 hidden layers.

The metrics for experimental evaluations were
accuracy and macro F1. The macro F1 was calculated by averaging F1 scores for respective classes, which biases towards rare classes compared with accuracy.
In practical TTS applications, the evaluation of Macro F1 is important for correctly generating the intended prosody control.

\subsection{Evaluation of the combination of acoustic and linguistic models}

To evaluate the effectiveness of combining acoustic and linguistic features, we compared multiple combinations.
Specifically, we used five acoustic inputs: HuBERT-base, Whisper-small, melspectrogram (Melspec), F0, and no acoustic input.
F0 contours were extracted by DIO \cite{morise2016world}.
We employed four linguistic inputs: PnG BERT, PL-BERT, one-hot phoneme sequence, and no linguistic input.

Table~\ref{table:combination} shows the results.
When HuBERT-base was used as the acoustic model,
PnG BERT had the highest accuracy and macro F1 scores for ACC, HL, and BI labels.
Although the scores of PL-BERT were relatively lower than those of PnG BERT, they tended to be higher than those of the one-hot phoneme input.
Hence, it is evident that the pretrained linguistic foundation models were effective in improving label prediction.
A similar tendency was shown when Whisper-small was used as the acoustic model.

When PnG BERT was used as the linguistic model and we compared acoustic feature inputs,
HuBERT-base and Whisper-small gave higher scores than melspectrogram and F0.
The same ranking was seen when we compared acoustic feature inputs without using linguistic inputs.
This indicates that speech foundation models are also effective in automatic prosody annotation.

When we used PnG BERT with no acoustic model,
we saw 82.5\% accuracy in the ACC label prediction.
Since this case can be regarded as the case of speech synthesis, where the speech waveform is not input,
it shows that PnG BERT has a certain ability for prosody prediction in TTS.

\subsection{Detail exploration of the combination of linguistic and acoustic models}

To explore the effectiveness of HuBERT and PnG BERT in prosody prediction, we illustrate confusion matrices of respective labels when we employ three input cases: ``HuBERT \& PnG BERT,'' ``HuBERT \& no linguistic input,'' and ``No acoustic input \& PnG BERT'' in Figs.~\ref{fig:confusion_acc} to \ref{fig:confusion_pau}.

From the results of ACC prediction, the use of PnG BERT enhanced overall prediction accuracy compared with no linguistic input. When we used no acoustic input, boundary pitch movement symbols (``\#'', ``\%'', ``?'') had more errors than when using HuBERT. This is likely because the boundary pitch movements are less dependent on input texts and are represented acoustically.

In BI prediction, when comparing the use of PnG BERT and no linguistic input, we found that the prediction accuracy between ``0'' and ``1'' was improved by incorporating PnG BERT.
This improvement can be attributed to the use of a tokenizer that can determine word boundaries. The prediction of ``F'' and ``D'' was also improved by PnG BERT. A possible reason is that the words corresponding to filled pauses and disfluency could be inferred by the linguistic model.
Comparing HuBERT with no acoustic input, it is seen that the confusion among ``1'', ``2'', and ``3'' were reduced by using HuBERT. This is due to the incorporation of acoustic features that affect prosodic phrase boundaries.

In terms of PAU prediction, the case with no acoustic input was unable to distinguish the presence of short pauses.
On the other hand, the introduction of HuBERT could extract the pause presence.

\subsection{Evaluation of acoustic models}
\label{sect:aco_feat}

To evaluate the effect of speech foundation models on prosody prediction, we compared diverse SSL models and Whisper sizes.
In these experimental evaluations, we used PnG BERT as the linguistic model. 
Tables~\ref{table:ssl_models} and \ref{table:whisper} show the prediction scores using SSL models and Whisper models, respectively.
When we compared SSL models including multiple sizes and training languages, we found that the models trained with a Japanese database yielded higher scores than those trained with English and multiple language databases.
Among Whisper models,
Whisper-large-v3 gave the highest prediction accuracy and F1 scores.

\begin{figure}[t]
  \centering
  \includegraphics[width=0.8\linewidth]{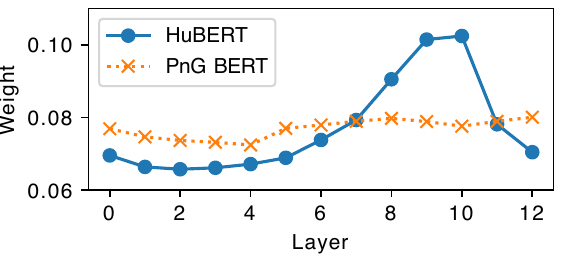}
  \caption{Layer weights for the weighted sum for HUBERT and PnG BERT.}
  \label{fig:layer_weight}
\end{figure}

\subsection{Analysis of the proposed model}

Finally, we illustrate the predicted prosody labels using the combination of HuBERT and PnG BERT in Fig.~\ref{fig:predicted_labels}.
We observed that high-low (HL) prediction results were consistent between manual and automatic annotations except for the last three moras.
The BI prediction errors were found between the labels ``1'' and ``2''.
Overall, we see that the automatic prosody annotation could predict the labels similar to those of manual annotation.

Figure~\ref{fig:layer_weight} shows the trained layer weights for the weighted sum of respective layers of SSL models when we used the combination of HuBERT and PnG BERT.
We find that all layers of PnG BERT were uniformly used and the 8--10 layers of HuBERT had larger weights than the other layers.

\section{Conclusions}

In this paper, we proposed an automatic prosody annotation model that uses the encoder hidden layers of both speech and linguistic foundation models.
The experimental results showed that the combination of speech and linguistic models enhanced the prosody label prediction accuracies compared to using either acoustic or linguistic inputs alone.
We also analyzed the types of prediction errors by confusion matrices.

Since the purpose of automatic prosody annotation is for speech synthesis,
in future work, we will investigate the effectiveness of the predicted labels on the training of speech synthesis. 
Additionally, because this paper performed experiments only using Japanese prosody labels, the effect on other languages should  be examined.

\clearpage

\bibliographystyle{IEEEtran}
\bibliography{mybib}

\begin{thebibliography}{10}
\providecommand{\url}[1]{#1}
\csname url@samestyle\endcsname
\providecommand{\newblock}{\relax}
\providecommand{\bibinfo}[2]{#2}
\providecommand{\BIBentrySTDinterwordspacing}{\spaceskip=0pt\relax}
\providecommand{\BIBentryALTinterwordstretchfactor}{4}
\providecommand{\BIBentryALTinterwordspacing}{\spaceskip=\fontdimen2\font plus
\BIBentryALTinterwordstretchfactor\fontdimen3\font minus \fontdimen4\font\relax}
\providecommand{\BIBforeignlanguage}[2]{{%
\expandafter\ifx\csname l@#1\endcsname\relax
\typeout{** WARNING: IEEEtran.bst: No hyphenation pattern has been}%
\typeout{** loaded for the language `#1'. Using the pattern for}%
\typeout{** the default language instead.}%
\else
\language=\csname l@#1\endcsname
\fi
#2}}
\providecommand{\BIBdecl}{\relax}
\BIBdecl

\bibitem{ogura2025mora}
T.~Ogura, T.~Okamoto, Y.~Ohtani, E.~Cooper, T.~Toda, and H.~Kawai, ``Mora-level prosody prediction for text-to-speech using japanese {BERT} without accentual labels,'' in \emph{Proc. ICASSP}, 2025.

\bibitem{koriyama2019semisupervised}
T.~Koriyama and T.~Kobayashi, ``Semi-supervised prosody modeling using deep {Gaussian} process latent variable model,'' in \emph{Proc. Interspeech}, 2019, pp. 4450--4454.

\bibitem{yufune2021accent}
K.~Yufune, T.~Koriyama, S.~Takamichi, and H.~Saruwatari, ``Accent modeling of low-resourced dialect in pitch accent language using variational autoencoder,'' in \emph{Proc. 11th ISCA Speech Synthesis Workshop (SSW 11)}, 2021, pp. 189--194.

\bibitem{yamauchi2024cross}
K.~Yamauchi, Y.~Saito, and H.~Saruwatari, ``Cross-dialect text-to-speech in pitch-accent language incorporating multi-dialect phoneme-level {BERT},'' in \emph{Proc. IEEE Spoken Language Technology Workshop (SLT)}, 2024, pp. 750--757.

\bibitem{zhong2024multimodal}
J.~Zhong, Y.~Li, H.~Huang, K.~Richmond, J.~Liu, Z.~Su, J.~Guo, B.~Tang, and F.~Zhu, ``Multi-modal automatic prosody annotation with contrastive pretraining of speech-silence and word-punctuation,'' in \emph{Proc. Interspeech}, 2024, pp. 2305--2309.

\bibitem{liu2024language}
C.~Liu, Z.-H. Ling, and Y.-J. Hu, ``Language-independent prosody-enhanced speech representations for multilingual speech synthesis,'' in \emph{Proc. IEEE Spoken Language Technology Workshop (SLT)}, 2024, pp. 482--488.

\bibitem{dai2022automatic}
Z.~Dai, J.~Yu, Y.~Wang, N.~Chen, Y.~Bian, G.~Li, D.~Cai, and D.~Yu, ``Automatic prosody annotation with pre-trained text-speech model,'' in \emph{Proc. Interspeech}, 2022, pp. 5513--5517.

\bibitem{shirahata2024audio}
Y.~Shirahata, B.~Park, R.~Yamamoto, and K.~Tachibana, ``Audio-conditioned phonemic and prosodic annotation for building text-to-speech models from unlabeled speech data,'' in \emph{Proc. Interspeech}, 2024, pp. 2795--2799.

\bibitem{kurihara2024low}
K.~Kurihara and M.~Sano, ``Low-resourced phonetic and prosodic feature estimation with self-supervised-learning-based acoustic modeling,'' in \emph{Proc. IEEE International Conference on Acoustics, Speech, and Signal Processing Workshops (ICASSPW)}, 2024, pp. 640--644.

\bibitem{vetter2019crosslingual}
M.~Vetter, S.~Sakti, and S.~Nakamura, ``Cross-lingual speech-based {ToBI} label generation using bidirectional {LSTM},'' in \emph{Proc. ICASSP}, 2019, pp. 6620--6624.

\bibitem{park2022unified}
B.~Park, R.~Yamamoto, and K.~Tachibana, ``A unified accent estimation method based on multi-task learning for {Japanese} text-to-speech,'' in \emph{Proc. Interspeech}, 2022, pp. 1931--1935.

\bibitem{zhai2023wav2tobi}
W.~Zhai and M.~Hasegawa-Johnson, ``{Wav2ToBI}: a new approach to automatic {ToBI} transcription,'' in \emph{Proc. Interspeech}, 2023, pp. 2748--2752.

\bibitem{gulati20conformer}
A.~Gulati, J.~Qin, C.-C. Chiu, N.~Parmar, Y.~Zhang, J.~Yu, W.~Han, S.~Wang, Z.~Zhang, Y.~Wu, and R.~Pang, ``Conformer: Convolution-augmented transformer for speech recognition,'' in \emph{Proc. Interspeech}, 2020, pp. 5036--5040.

\bibitem{devlin2019bert}
J.~Devlin, M.-W. Chang, K.~Lee, and K.~Toutanova, ``{BERT}: Pre-training of deep bidirectional transformers for language understanding,'' in \emph{Proc. NAACL}, 2019, pp. 4171--4186.

\bibitem{radford2023robust}
A.~Radford, J.~W. Kim, T.~Xu, G.~Brockman, C.~McLeavey, and I.~Sutskever, ``Robust speech recognition via large-scale weak supervision,'' in \emph{Proc. ICML}, 2023, pp. 28\,492--28\,518.

\bibitem{baevski2020wav2vec}
A.~Baevski, Y.~Zhou, A.~Mohamed, and M.~Auli, ``wav2vec 2.0: A framework for self-supervised learning of speech representations,'' \emph{Proc. NeurIPS}, vol.~33, pp. 12\,449--12\,460, 2020.

\bibitem{hayashi2019pretrained}
T.~Hayashi, S.~Watanabe, T.~Toda, K.~Takeda, S.~Toshniwal, and K.~Livescu, ``Pre-trained text embeddings for enhanced text-to-speech synthesis,'' in \emph{Proc. Interspeech}, 2019, pp. 4430--4434.

\bibitem{xiao2020improving}
Y.~Xiao, L.~He, H.~Ming, and F.~K. Soong, ``Improving prosody with linguistic and {BERT} derived features in multi-speaker based {Mandarin} {Chinese} neural {TTS},'' in \emph{Proc. ICASSP}, 2020, pp. 6704--6708.

\bibitem{kenter2020improving}
T.~Kenter, M.~Sharma, and R.~Clark, ``Improving the prosody of {RNN}-based {English} text-to-speech synthesis by incorporating a {BERT} model,'' in \emph{Proc. Interspeech}, 2020, pp. 4412--4416.

\bibitem{xu2021improving}
G.~Xu, W.~Song, Z.~Zhang, C.~Zhang, X.~He, and B.~Zhou, ``Improving prosody modelling with cross-utterance {BERT} embeddings for end-to-end speech synthesis,'' in \emph{Proc. ICASSP}, 2021, pp. 6079--6083.

\bibitem{yasuda2019investigation}
Y.~{Yasuda}, X.~{Wang}, S.~{Takaki}, and J.~{Yamagishi}, ``Investigation of enhanced {Tacotron} text-to-speech synthesis systems with self-attention for pitch accent language,'' in \emph{Proc. ICASSP}, 2019, pp. 6905--6909.

\bibitem{liu2024pewav2vec}
Z.-C. Liu, L.~Chen, Y.-J. Hu, Z.-H. Ling, and J.~Pan, ``{PE-Wav2vec}: A prosody-enhanced speech model for self-supervised prosody learning in {TTS},'' \emph{IEEE/ACM Transactions on Audio, Speech, and Language Processing}, vol.~32, pp. 4199--4210, 2024.

\bibitem{jia2021png}
Y.~Jia, H.~Zen, J.~Shen, Y.~Zhang, and Y.~Wu, ``{PnG BERT}: Augmented {BERT} on phonemes and graphemes for neural {TTS},'' in \emph{Proc. Interspeech}, 2021, pp. 151--155.

\bibitem{li2023phoneme}
Y.~A. Li, C.~Han, X.~Jiang, and N.~Mesgarani, ``Phoneme-level {BERT} for enhanced prosody of text-to-speech with grapheme predictions,'' \emph{arXiv preprint arXiv:2301.08810}, 2023.

\bibitem{kunesova2022detection}
M.~Kunešová and M.~Řezáčková, ``Detection of prosodic boundaries in speech using wav2vec 2.0,'' in \emph{Proc. International Conference on Text, Speech, and Dialogue}, 2022, p. 377–388.

\bibitem{Maekawa_2003}
K.~Maekawa, ``{Corpus of Spontaneous Japanese: Its design and evaluation},'' in \emph{ISCA \& IEEE Workshop on Spontaneous Speech Processing and Recognition}, 2003.

\bibitem{hsu2021hubert}
W.-N. Hsu, B.~Bolte, Y.-H.~H. Tsai, K.~Lakhotia, R.~Salakhutdinov, and A.~Mohamed, ``{HuBERT}: Self-supervised speech representation learning by masked prediction of hidden units,'' \emph{IEEE/ACM Transactions on Audio, Speech, and Language Processing}, vol.~29, pp. 3451--3460, 2021.

\bibitem{chen2022wavlm}
S.~Chen, C.~Wang, Z.~Chen, Y.~Wu, S.~Liu, Z.~Chen, J.~Li, N.~Kanda, T.~Yoshioka, X.~Xiao, J.~Wu, L.~Zhou, S.~Ren, Y.~Qian, Y.~Qian, J.~Wu, M.~Zeng, X.~Yu, and F.~Wei, ``{WavLM}: Large-scale self-supervised pre-training for full stack speech processing,'' \emph{IEEE Journal of Selected Topics in Signal Processing}, vol.~16, no.~6, pp. 1505--1518, 2022.

\bibitem{mohamed2022self}
A.~Mohamed, H.-y. Lee, L.~Borgholt, J.~D. Havtorn, J.~Edin, C.~Igel, K.~Kirchhoff, S.-W. Li, K.~Livescu, L.~Maal{\o}e \emph{et~al.}, ``Self-supervised speech representation learning: A review,'' \emph{IEEE Journal of Selected Topics in Signal Processing}, vol.~16, no.~6, pp. 1179--1210, 2022.

\bibitem{li2023styletts2}
Y.~A. Li, C.~Han, V.~Raghavan, G.~Mischler, and N.~Mesgarani, ``{StyleTTS} 2: Towards human-level text-to-speech through style diffusion and adversarial training with large speech language models,'' in \emph{Proc. NeurIPS}, vol.~36, 2023, pp. 19\,594--19\,621.

\bibitem{ma2024emobox}
Z.~Ma, M.~Chen, H.~Zhang, Z.~Zheng, W.~Chen, X.~Li, J.~Ye, X.~Chen, and T.~Hain, ``Emobox: Multilingual multi-corpus speech emotion recognition toolkit and benchmark,'' in \emph{Proc. Interspeech}, 2024, pp. 1580--1584.

\bibitem{mogridge2024nonintrusive}
R.~Mogridge, G.~Close, R.~Sutherland, T.~Hain, J.~Barker, S.~Goetze, and A.~Ragni, ``Non-intrusive speech intelligibility prediction for hearing-impaired users using intermediate {ASR} features and human memory models,'' in \emph{Proc. ICASSP}, 2024, pp. 306--310.

\bibitem{mcauliffe2017montreal}
M.~McAuliffe, M.~Socolof, S.~Mihuc, M.~Wagner, and M.~Sonderegger, ``Montreal forced aligner: Trainable text-speech alignment using kaldi.'' in \emph{Proc. Interspeech}, vol. 2017, 2017, pp. 498--502.

\bibitem{badlani2022one}
R.~Badlani, A.~{\L}a{\'n}cucki, K.~J. Shih, R.~Valle, W.~Ping, and B.~Catanzaro, ``One {TTS} alignment to rule them all,'' in \emph{Proc. ICASSP}, 2022, pp. 6092--6096.

\bibitem{koriyama2025vae}
T.~Koriyama, ``{VAE}-based phoneme alignment using gradient annealing and {SSL} acoustic features,'' in \emph{Proc. Interspeech}, 2024, pp. 3814--3818.

\bibitem{peyasantiwong1986stress}
P.~Peyasantiwong, ``Stress in {Thai},'' in \emph{Papers from a Conference on Thai Studies in Honor of William J. Gedney. Michigan Papers on South and Southeast Asia, Center for South and Southeast Asian Studies, University of Michigan, Ann Arbor}, 1986, pp. 19--39.

\bibitem{wikipedia_corpus}
Japanese Wikipedia Corpus,\\ \url{https://dumps.wikimedia.org/jawiki/}.

\bibitem{cc100}
CC-100: Monolingual Datasets from Web Crawl Data,\\ \url{https://data.statmt.org/cc-100/}.

\bibitem{unidic}
UniDic, \\ \url{https://clrd.ninjal.ac.jp/unidic/}.

\bibitem{kingma2014adam}
D.~P. Kingma and J.~Ba, ``Adam: A method for stochastic optimization,'' in \emph{Proc. ICLR}, 2015.

\bibitem{kurihara2021prosodic}
K.~Kurihara, N.~Seimiya, and T.~Kumano, ``Prosodic features control by symbols as input of sequence-to-sequence acoustic modeling for neural {TTS},'' \emph{IEICE Transactions on Information and Systems}, vol. E104.D, no.~2, pp. 302--311, 2021.

\bibitem{Maekawa_2002}
K.~Maekawa, H.~Kikuchi, Y.~Igarashi, and J.~Venditti, ``{X-JToBI: an extended J-ToBI for spontaneous speech},'' in \emph{Proc. 7th ICSLP}, 2002, pp. 1545--1548.

\bibitem{watanabe2018espnet}
S.~Watanabe, T.~Hori, S.~Karita, T.~Hayashi, J.~Nishitoba, Y.~Unno, N.~Enrique Yalta~Soplin, J.~Heymann, M.~Wiesner, N.~Chen \emph{et~al.}, ``{ESPnet}: End-to-end speech processing toolkit,'' in \emph{Proc. Interspeech}, 2018.

\bibitem{reazonspeech}
ReazonSpeech \\ \url{https://research.reazon.jp/projects/ReazonSpeech/}.

\bibitem{rinna_hubert}
Rinna/japanese-hubert-base \\ \url{https://huggingface.co/rinna/japanese-hubert-base/}.

\bibitem{morise2016world}
M.~Morise, F.~Yokomori, and K.~Ozawa, ``{WORLD}: A vocoder-based high-quality speech synthesis system for real-time applications,'' \emph{IEICE Transactions on Information and Systems}, vol. E99.D, no.~7, pp. 1877--1884, 2016.

\end{thebibliography}

\end{document}